\documentclass[prc,twocolumn,showpacs]{revtex4}
\usepackage{graphicx}
\usepackage{epsfig}
\usepackage{bm}

\begin{document}

\title{Suppression of complete fusion due to breakup in the reactions
$^{10,11}$B + $^{209}$Bi}

\author{L.R. Gasques\footnote[1]{ Current address: Centro de F\'{\i}sica
Nuclear da Universidade de Lisboa, Lisboa, Portugal.}, D.J. Hinde,
M. Dasgupta, A. Mukherjee\footnote[2]{ Permanent Address: Saha Institute
of Nuclear Physics, 1/AF, Bidhan Nagar, Kolkata-700064, India} and R.G.
Thomas\footnote[3]{ Current address: B.A.R.C., Mumbai, India.}}
\affiliation{Department of Nuclear Physics, Research School of Physical
Sciences and Engineering, Australian National University, Canberra, ACT 0200,
Australia}

\date{\today}
\begin{abstract}
Above-barrier cross sections of  $\alpha$-active heavy reaction
products, as well as fission, were measured for the reactions of
$^{10,11}$B with $^{209}$Bi. Detailed analysis showed that the heavy products
include components from incomplete fusion as well as complete fusion (CF),
but fission originates almost exclusively from CF.
Compared with fusion calculations without breakup, the CF cross sections are
suppressed by 15\%
for $^{10}$B and 7\% for $^{11}$B. A consistent and systematic variation of
the
suppression of CF for reactions of the weakly bound nuclei $^{6,7}$Li,
$^{9}$Be, $^{10,11}$B on targets of $^{208}$Pb and $^{209}$Bi is found as
a function of the breakup threshold energy.
\end{abstract}


\maketitle

%%%%%%%%%%%%%%%%%%%%%%%%%%%%%%%%%%%%%%%%%%%%%%%%%%%%%%%%%%%%%%%%%%%%%%%%%%%%%%%
\section{Introduction}
\label{introduct}
%%%%%%%%%%%%%%%%%%%%%%%%%%%%%%%%%%%%%%%%%%%%%%%%%%%%%%%%%%%%%%%%%%%%%%%%%%%%%%%

Over the last decade, the effect of the breakup of weakly bound nuclei on
both fusion and other
reaction processes has been widely investigated, from
theoretical and experimental perspectives~\cite{Can06}.  Studies
with weakly bound light stable nuclei indicate that complete fusion (CF)
cross sections (defined experimentally~\cite{Das99,Das04} as absorption of
all the charge of projectile) are suppressed at above-barrier
energies in comparison with the predictions of both the single
barrier penetration model (SBPM) and the coupled-channels (CC)
model~\cite{Das99,Das04,Gom06,Tri02,Muk06,Wu03}. This is attributed to the
low binding energy of the projectiles, which can break up prior to
reaching the fusion barrier. In general, the missing CF cross sections are
found in yields of incomplete fusion (ICF), which occurs when not
all the fragments are captured by the target.

Calculation of CF cross sections, as distinct from ICF, is currently
not possible using quantum-mechanical scattering theories such as
the continuum discretised coupled-channels model (CDCC), which can
only predict total (CF+ICF) fusion. To address this problem, a
three-dimensional classical dynamical model was recently
developed~\cite{Tor07}, allowing the calculation of CF and ICF cross sections
at energies above the fusion barrier. The model relates
above-barrier CF and ICF yields with below-barrier no-capture
breakup, where the Coulomb barrier inhibits capture of charged
fragments by the target nucleus. This is achieved by the
introduction of a stochastically sampled breakup function, which can
be obtained experimentally from the measurement of no-capture
breakup cross sections at sub-barrier energies~\cite{Hin02}, or from
CDCC calculations. A quantitative relationship between the CF and
ICF at energies above the barrier and the below-barrier no-capture
breakup opens up a new approach to studying the influence of breakup
on fusion~\cite{Hin02}. To test and exploit this new approach, both
below-barrier and above-barrier measurements are required, for light
nuclei with a range of breakup threshold energies. Ideally, all
significantly populated CF and ICF channels should be measured, to
allow detailed comparison with the model predictions.

As part of this development, the present paper describes
measurements of CF and ICF cross sections for the
$^{10,11}$B~+~$^{209}$Bi reactions at energies above the fusion
barrier V$_{B}$, ranging from 1.1V$_{B}$ to 1.5V$_{B}$. A comparison
of the above-barrier CF cross sections with SBPM calculations of
fusion without breakup allowed the determination of the suppression
of CF due to breakup of the $^{10}$B and $^{11}$B
projectiles~\cite{Das04}. The most favorable breakup thresholds are quite
different, being 4.461 MeV for $^{10}$B and 8.665 MeV for $^{11}$B.
The CF suppression is compared with those obtained for reactions
involving other stable but weakly bound projectiles.

%%%%%%%%%%%%%%%%%%%%%%%%%%%%%%%%%%%%%%%%%%%%%%%%%%%%%%%%%%%%%%%%%%%%%%%%%%%%%%
\section{Experimental Method}
\label{method}
%%%%%%%%%%%%%%%%%%%%%%%%%%%%%%%%%%%%%%%%%%%%%%%%%%%%%%%%%%%%%%%%%%%%%%%%%%%%%%

The experiments were performed using pulsed $^{10,11}$B beams from
the 14UD tandem electrostatic accelerator at the Australian National
University. The beams, with a pulse width of 1~ns and interval of
640 ns, were incident on self-supporting $^{nat}$Bi targets of
thickness 480 $\mu$g cm$^{-2}$. Fission measurements were made with
an unbacked target, whilst the yield of the heavy CF and ICF products were
measured by their $\alpha$-decay, using four Bi targets backed by
560 $\mu$g cm$^{-2}$ Al foils to stop the recoiling evaporation
residues (ER). The beam species, beam energy and target cycling sequence
were chosen to minimize the buildup of activity contaminating
subsequent measurements. The experimental arrangement was similar to that
reported in Ref.~\cite{Das04}, and only a brief description is given
here.  Fission events following fusion were measured using two
large area position sensitive multiwire proportional counters~\cite{Ber01}.
Each detector, with an active area of 28.4 $\times$
35.7 cm$^2$, was located 18.0 cm from the target. To identify
the heavy reaction products, their characteristic
decay $\alpha$-particle energies were used,
and cross sections were determined using their known $\alpha$
branching ratios and half lives. The $\alpha$-particles from
short-lived activities (T$_{1/2} \leq$ 24 min) were detected between
the beam pulses by a back-angle annular silicon surface barrier detector.
Alpha decays from long-lived activities were measured using a
silicon surface barrier detector placed below the annular counter,
close to the plane of the target ladder~\cite{Das04}. For normalization,
two Si surface barrier detectors
(monitors), symmetrically placed about the beam axis at
22.5$^{\circ}$, were used to measure elastically scattered beam
particles. Absolute cross sections for
fission and ER were determined by performing calibrations at
a sub-barrier energy in which elastically scattered $^{58}$Ni
projectiles of 120 MeV were detected in the fission detectors and
the annular $\alpha$-counter as well as in the two forward-angle
monitors. The relative solid angles of the annular and close
geometry Si detectors was determined using the 24 minute
$\alpha$-activity from $^{212}$Rn~\cite{Das04}.

\begin{figure}[t]
\begin{center}
\epsfig{file=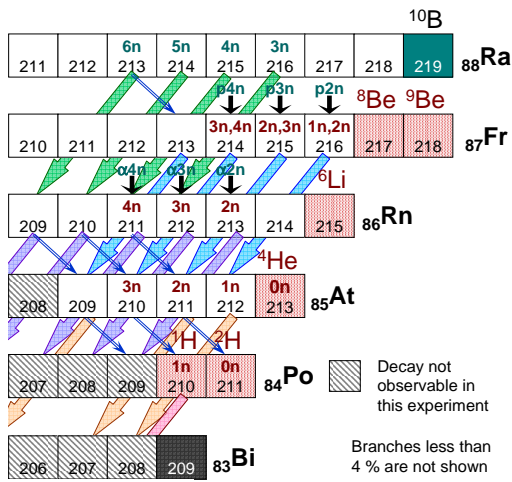,angle=-90,clip=,width=0.8\linewidth}
\caption{(color online). Diagram showing the significant production and decay
paths of the isotopes measured for the $^{10}$B + $^{209}$Bi reaction,
including production both by CF and ICF. The nucleus captured is shown above
the shaded square corresponding to the
compound nucleus (CN) on the right. For the isotopes observed, the evaporation
channel is labeled assuming only neutron evaporation from each CN.
Possible production by charged particle evaporation after CF forming
$^{219}$Ra is indicated by the labeled vertical black arrows.
$\alpha$-particle decays are indicated
by the large shaded diagonal arrows, which lead from the parent to the
daughter nucleus.
Electron capture decay is shown by the narrow  diagonal arrows.
The assignment of
cross sections accounted for all these population and decay pathways.}
\label{production}
\end{center}
\end{figure}

%%%%%%%%%%%%%%%%%%%%%%%%%%%%%%%%%%%%%%%%%%%%%%%%%%%%%%%%%%%%%%%%%%%%%%%%%%%%%%
\section{Determination of Complete Fusion Yields}
\label{CF}
%%%%%%%%%%%%%%%%%%%%%%%%%%%%%%%%%%%%%%%%%%%%%%%%%%%%%%%%%%%%%%%%%%%%%%%%%%%%%%

The identification and determination of the yields of the complete
fusion products, amongst the many possible reaction products, will
be described in order of the complexity of the procedure. Firstly
the $xn$ evaporation products following CF will be discussed, then
fission following CF, and finally the products of $pxn$ and
$\alpha$$xn$ evaporation following CF. To illustrate the complexity
of the analysis, the many possible paths for the production of all
$\alpha$-active products observed for the $^{10}$B-induced reaction
are shown in Fig.~\ref{production}.

\subsection{Evaporation products following complete fusion}

\begin{figure}[!]
\begin{center}
\epsfig{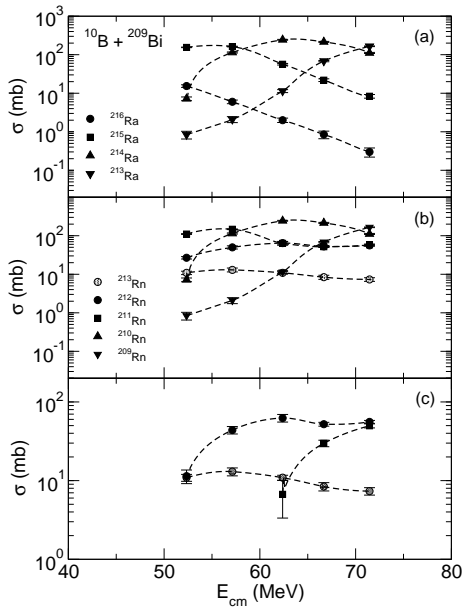}
\caption{(a) The measured  cross sections for the production
of Ra isotopes, (b) of Rn isotopes, and (c) yield of Rn isotopes after 
subtracting the contributions from the decay of Ra isotopes for the reaction 
$^{10}$B~+~$^{209}$Bi. The lines guide the eye.}
\label{10b-xn}
\end{center}
\end{figure}

\begin{figure}[!]
\begin{center}
\epsfig{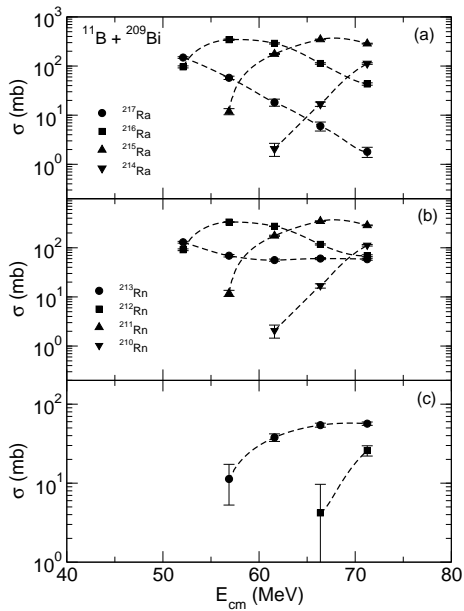}
\caption{(a) The measured cross sections for the production
of Ra isotopes, (b) of Rn isotopes, and (c) yield of Rn isotopes after 
subtracting the contributions from the decay of
Ra isotopes for the reaction $^{11}$B~+~$^{209}$Bi. The lines
guide the eye.}
\label{11b-xn}
\end{center}
\end{figure}

Complete fusion of $^{10,11}$B incident on $^{209}$Bi leads to the
compound nuclei $^{219,220}$Ra, respectively.  Neutron evaporation is
the dominant evaporation mode, producing residual Ra evaporation residues,
all of which are $\alpha$-active (the number of neutrons
evaporated in the formation of the observed isotopes following $^{10}$B
fusion is
indicated in Fig.~\ref{production}, for example for $^{214}$Ra by
the legend $5n$).
Since they can only be produced following complete fusion,
interpretation of their measured cross sections is unambiguous.
They are shown in Figs.~\ref{10b-xn}(a) and~\ref{11b-xn}(a)
for the reactions $^{10,11}$B~+~$^{209}$Bi, respectively. As
expected, each evaporation channel shows a gradual rise and fall
with increasing energy, in accord with expectations from the neutron binding
energies and average neutron kinetic energies.

The $\alpha$-decay of these Ra nuclei
produces Rn daughter nuclei, which also undergo $\alpha$-decay, in turn
forming Po
daughter nuclei. Decay to daughter nuclei is indicated
in Fig.~\ref{production} by diagonal
shaded arrows. The measured cross sections for the Rn
isotopes are presented in Figs.~\ref{10b-xn}(b) and~\ref{11b-xn}(b).
For both reactions, they can be significantly larger than those
predicted from the parent Ra yields, indicating that there are other
mechanisms directly populating the Rn nuclei, as might be expected.
Subtracting the expected Rn cross sections resulting from the
Ra parent decay, the cross sections for the direct
production of Rn isotopes are shown for $^{10}$B in
Fig.~\ref{10b-xn}(c), and for $^{11}$B in Fig.~\ref{11b-xn}(c).
The $\alpha$ energy spectra also showed the presence of small cross
sections for production of Fr isotopes (not plotted).
The origin of Fr and Rn isotopes are discussed next.

\subsection{Production mechanisms of Fr, Rn products}

The direct production yields of Fr and Rn can result from CF, through $pxn$
and $\alpha$$xn$ evaporation respectively, in competition with
$xn$ evaporation and fission. These pathways are
indicated in Fig.~\ref{production} by the vertical arrows, identified with the
specific evaporation channel leading to the observed isotope
(e.g. $p4n$ leading to $^{214}$Fr). However, this is not the only mechanism 
available. The $^{10,11}$B projectiles can break up before fusion,
into a number of different mass partitions. The most energetically favorable
$^{10}$B breakup channel is
\begin{eqnarray*}
^{10}\mbox{B} &\rightarrow & ^{6}\mbox{Li} \mbox{+}  ^4\mbox{He};
~~~~~~~ \mbox{Q = --4.461 MeV}.
\end{eqnarray*}
Less favored break up channels are:
\begin{eqnarray*}
^{10}\mbox{B} &\rightarrow &  ^{8}\mbox{Be} \mbox{+}  ^2\mbox{H};~~~~~~~~
\mbox{Q = --6.026 MeV}\\
   & \rightarrow           & ^{9}\mbox{Be} \mbox{+}  ^{1}\mbox{H};~~~~~~~~
\mbox{Q = --6.586~MeV}\\
   & \rightarrow &  ^4\mbox{He} \mbox{+} ^4\mbox{He} \mbox{+}  ^{2}\mbox{H};~
\mbox{Q = --5.934~MeV}.
\end{eqnarray*}
Similarly the most favorable $^{11}$B break up channel is:
\begin{eqnarray*}
^{11}\mbox{B} &\rightarrow & ^{7}\mbox{Li} \mbox{+} ^4\mbox{He};
~~~~~~~ \mbox{Q = --8.665 MeV}.
\end{eqnarray*}
It may also break up into:
\begin{eqnarray*}
~ ^{11}\mbox{B} &\rightarrow &  ^{8}\mbox{Be} \mbox{+} ^3\mbox{H};~~~~~~~~
\mbox{Q = --11.224 MeV}\\
   & \rightarrow &  ^4\mbox{He} \mbox{+} ^4\mbox{He} \mbox{+} ^{3}\mbox{H};~
\mbox{Q = --11.132~MeV}.
\end{eqnarray*}

The capture and absorption of one of these fragments by $^{209}$Bi
would form one of the compound nuclei $^{217,218}$Fr, $^{215,216}$Rn,
$^{213}$At or $^{210,211}$Po.
The CN formed by ICF in the $^{10}$B reaction are indicated in
Fig.~\ref{production} by shading, with the
specific captured fragment noted (e.g. $^{6}$Li).
Isotopes of Fr and Rn would be produced, as well as isotopes of At and Po,
after evaporation of neutrons (the dominant evaporation channel)
from these compound nuclei; these are indicated in Fig.~\ref{production} by the
number of neutrons evaporated, for those isotopes that were observed in the
experiment.

Because of the different pathways available in the $^{10,11}$B~+~$^{209}$Bi
reactions, the yields of Rn isotopes originating from $\alpha$$xn$
evaporation following
complete fusion cannot be directly separated
from $xn$ evaporation following ICF with capture of a $^{6,7}$Li.
Equivalently, yields of Fr
isotopes produced after $pxn$ evaporation from the
CF compound nuclei $^{215,216}$Ra will be mixed with the same products formed
following capture of a $^{8,9}$Be fragment (or two $^{4}$He).
The resolution of this problem requires the introduction of more experimental
information, which will be described after the attribution
of the observed fission cross sections is discussed.

\begin{figure}[!]
\begin{center}
\includegraphics[width=8.00cm]{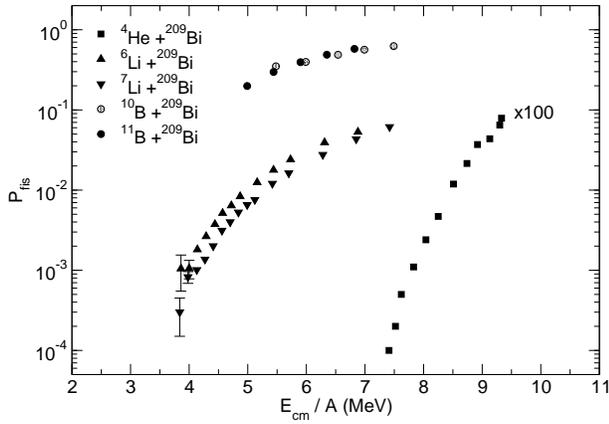}
\caption{Measured empirical fission probability as a function of the
energy per nucleon of the projectile for the
reactions $^{10,11}$B~+~$^{209}$Bi, $^{6,7}$Li~+~$^{209}$Bi and
$^{4}$He~+~$^{209}$Bi.}
\label{Pfis}
\end{center}
\end{figure}

\begin{figure}[!]
\begin{center}
\epsfig{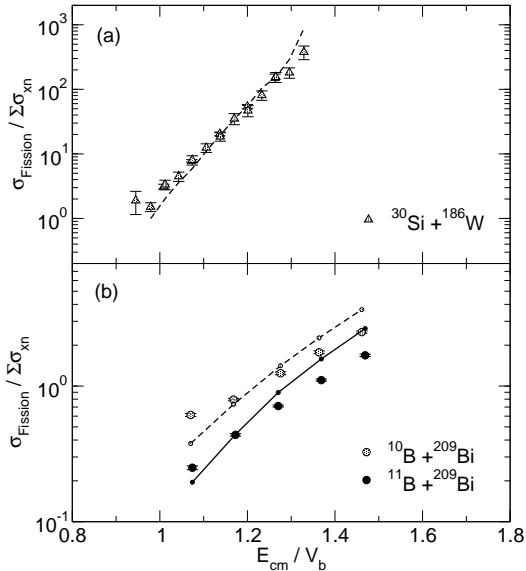} \caption{Ratio of the measured
 fission cross sections to the summed $xn$ cross sections for 
(a) $^{30}$Si~+~$^{186}$W and (b) $^{10,11}$B~+~$^{209}$Bi. The lines are the 
predictions of statistical model calculations (see text).} \label{ffis}
\end{center}
\end{figure}

\subsection{Attribution of Fission to Complete Fusion}

To investigate whether the contribution of fission following ICF is
significant, we define the empirical probability of fission following fusion  as
\begin{equation}
\label{Pfission}
P_{fis} = \frac {\sigma_{fis}}{\sigma_{fis}+\sigma_{CF}+\sigma_{ICF}},
\end{equation}
in which the determination of the observed ICF cross sections
($\sigma_{ICF}$) will be discussed in the next section. Experimental
values of $P_{fis}$ are plotted in Fig.~\ref{Pfis} as a function of
the bombarding energy per nucleon $(E/A)$, for the
reactions $^{10,11}$B~+~$^{209}$Bi, $^{6,7}$Li~+~$^{209}$Bi~\cite{Das04}
and $^{4}$He~+~$^{209}$Bi~\cite{Hui62}. The latter correspond to the incomplete fusion
reactions associated with the most favorable breakup channels. It is clear
from the large reduction in $P_{fis}$ as the captured mass decreases
that the contribution of fission following ICF of $^{6,7}$Li or
$^{4}$He will be negligible compared with that resulting from CF. As
an example, we can estimate the maximum contribution to the fission
yield from ICF for the $^{10}$B~+~$^{209}$Bi reaction. From Fig.~\ref{Pfis},
the maximum probability of fission after a $^{6}$Li
fragment is captured is 7\% at the highest value of $E/A$ (7.5 MeV).
It will be found in the next section that approximately 8\% of the total fusion
(CF+ICF) cross section is attributed to ICF with a $^{6}$Li
fragment, thus the fraction of the total fission is 0.07x0.08
divided by the fraction of CF resulting in fission at this energy
($\sim$0.6), giving less than 1$\%$ of the fission yield resulting from
ICF of $^{6}$Li. At lower $E/A$, the fraction is even smaller.
For the same $E/A$, the fission contribution
following the capture of a $^4$He fragment is very much lower, as
clearly seen in Fig.~\ref{Pfis}. These calculations are supported
by the measured folding angle distributions for fission, which were
consistent with complete fusion within experimental uncertainties.

The relatively small contribution
of fission following ICF can be understood qualitatively, since the
angular momentum and excitation energy brought in by an ICF fragment
is on average lower than when the entire projectile fuses with the target, and
the compound nucleus itself is less fissile~\cite{Das04}. Despite
the heavier projectile, the above quantitative analysis justifies using
the same approach as in Ref.~\cite{Das04}, and thus the measured
fission cross sections are attributed only to CF.

\begin{figure}[!]
\begin{center}
\epsfig{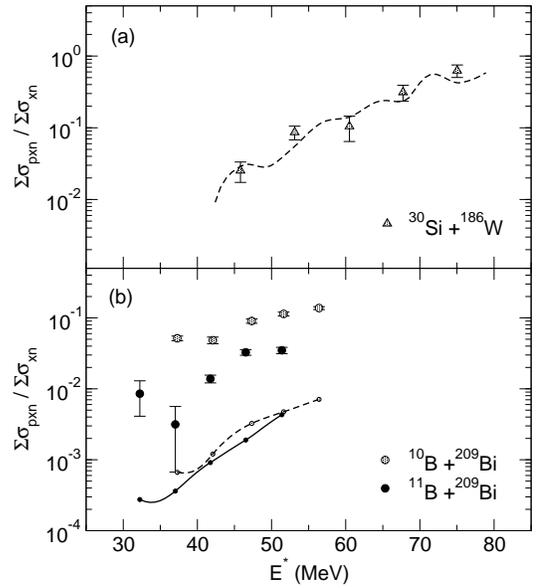}
\caption{Ratio of the summed $pxn$ to $xn$ cross sections for (a) 
$^{30}$Si~+~$^{186}$W
and (b) $^{10,11}$B~+~$^{209}$Bi. The lines are the predictions of
statistical model calculations.}
\label{fpxn}
\end{center}
\end{figure}

\subsection{Separation of Complete and Incomplete Fusion for Fr and Rn
Products}

\begin{figure}[!]
\begin{center}
\epsfig{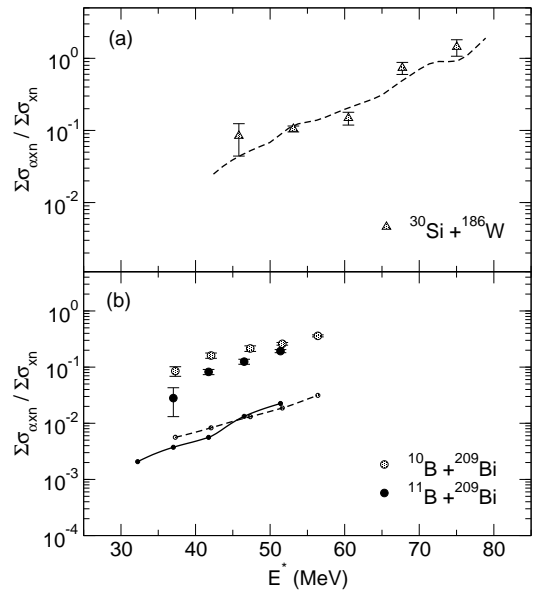}
\caption{Summed $\alpha$$xn$ to the $xn$ cross sections for (a)
$^{30}$Si~+~$^{186}$W and (b) $^{10,11}$B~+~$^{209}$Bi. The lines are the
predictions of statistical model calculations (see text).}
\label{faxn}
\end{center}
\end{figure}

The origin of the yields of Fr and Rn isotopes was determined making
use of measurements of above-barrier fusion cross sections for the
$^{30}$Si~+~$^{186}$W reaction~\cite{Ber01}, which forms the
$^{216}$Ra compound nucleus following fusion. The direct production
of Fr and Rn isotopes by ICF for this reaction should be
insignificant at all measured energies, as $^{30}$Si is expected to
behave as a normal, strongly bound, non-cluster structure nucleus.
Thus statistical model calculations~\cite{STAT} which reproduce both
the fission probabilities and the relative yields of $pxn$ and
$\alpha$$xn$ evaporation for the $^{30}$Si~+~$^{186}$W reaction
should reliably predict the charged particle evaporation for the
$^{10,11}$B reactions. The experimental ratios of fission cross
sections to $xn$ evaporation residues (plotted in this way to
minimise bias due to ICF) are shown for the three reactions in
Fig.~\ref{ffis}. The results for $^{30}$Si~+~$^{186}$W are
corrected for the inferred quasi-fission contribution~\cite{Ber01}.
It is necessary that the statistical model calculations reproduce
these experimental ratios to ensure that the ER calculations are for
the correct angular momentum distributions. The calculations, using
the code JOANNE2~\cite{STAT}, were matched to experiment by scaling
the fission barrier heights by factors 0.77, 0.83 and 0.85 for the
reactions leading to  $^{216}$Ra, $^{219}$Ra and $^{220}$Ra
respectively. A similar trend of barrier scaling factors with
neutron number has been found previously~\cite{Hin83}.
Figs.~\ref{fpxn}(a) and~\ref{faxn}(a) show the measured
ratios~\cite{Ber01thesis} of the cross sections for $pxn$ and
$\alpha$$xn$ products divided by the $xn$ cross sections, for the
calibration reaction $^{30}$Si~+~$^{186}$W. The JOANNE2 calculations
were found to give good agreement with the experimental data after
multiplying by a factor of 1.8 for $pxn$ and 1.5 for $\alpha$$xn$.
Using these scaling factors, the predicted $pxn$ and $\alpha$$xn$
relative yields for the $^{10,11}$B reactions [Fig.~\ref{fpxn}(b)
and Fig.~\ref{faxn}(b)] lie far below the current measurements. This
clearly indicates that a large fraction of the direct production of
Fr and Rn isotopes is due to ICF. For the $^{10}$B~+~$^{209}$Bi
reaction, the data are consistent with 93$\pm$2\% of the Fr
and 90$\pm$2\% of the Rn yields resulting from ICF, whilst for the
$^{11}$B~+~$^{209}$Bi reaction, 85$\pm$4\% of the Fr and
85$\pm$3\% of the Rn yields result from ICF. Because the deduced
cross sections of Fr and Rn from CF make such a small contribution
to the total CF cross sections, the uncertainties in the Fr and Rn
CF fractions do not make a significant difference to the deduced CF
cross-sections.

\subsection{Complete Fusion Cross Sections}

The CF cross sections were determined by summing the yields for the
Ra isotopes, the small fraction  of the cross sections for
Fr and Rn isotopes associated with CF ($\sim$8\% for $^{10}$B and
$\sim$15\% for $^{11}$B induced reactions), and the full fission cross
sections. The total measured yields in each category, together with
the deduced CF cross sections are presented in Table~\ref{yields-1}
for $^{10}$B~+~$^{209}$Bi and Table~\ref{yields-2} for
$^{11}$B~+~$^{209}$Bi. The center-of-mass energies have been
corrected for energy loss in the target. The errors given in the
cross sections only reflect statistical uncertainties. The
excitation functions for CF, for both reactions, are shown as filled
circles in Fig.~\ref{CF-2panels}. Having determined the CF yields,
the determination of the suppression of CF by ICF is discussed
in the next section. 

The yields of other products of ICF, such as At and Po isotopes, together with 
transfer reactions, whose details do not affect the subsequent analysis, are 
presented and discussed in the Appendix.
The sum of these cross sections are shown in Fig.~\ref{CF-2panels} by open 
triangles (multiplied by a factor of 2 for clarity in the figure). The sum of 
the measured cross sections for Rn and Fr not associated with CF are shown by 
open circles. Adding these cross sections to the deduced CF cross sections 
gives the total cross sections for almost all reaction products heavier than 
the target (indicated by open squares).
These exceed the calculated total fusion cross sections as expected,
as the cross sections for At and Po must include contributions from
transfer reactions.

\begin{table} [t]
\caption{Fission, evaporation residues and  complete fusion
cross sections for the $^{10}$B~+~$^{209}$Bi reaction. The
center-of-mass energies $E_{c.m.}$ are corrected for energy losses
in the target. The quoted errors reflect only statistical
uncertainties. For the Rn and Fr isotopes only 7\% and 10\%
respectively of the total yields presented in that table contributes
to the total complete fusion cross sections (see text).}
\label{yields-1}
\begin{center}
\begin{tabular}{lllllll}
\hline
\hline
$E_{beam}$ \    & $E_{c.m.}$ \
& $\sigma_{fission}$  & $\sum$$\sigma_{Ra}$
\    & $\sum$$\sigma_{Rn}$ \ \   &
$\sum$$\sigma_{Fr}$ \ \   & $\sigma_{CF}$\\
(MeV) \  & (MeV) \   & ~(mb)  & ~(mb) \   & ~(mb) \ \   & ~(mb)
\ \  & (mb)\\
\hline
54.84 & 52.34 & 82$\pm$1 & 134$\pm$3 & 11.4$\pm$2.2 & 6.9$\pm$1.0
& 217$\pm$3\\
59.86 & 57.13 & 217$\pm$2 & 273$\pm$5 & 44$\pm$5 & 13$\pm$2
& 494$\pm$5\\
65.36 & 62.38 & 400$\pm$4 & 322$\pm$7 & 69$\pm$8 & 29$\pm$3
& 730$\pm$8\\
69.87 & 66.68 & 559$\pm$6 & 315$\pm$7 & 82$\pm$8 & 36$\pm$4
& 883$\pm$9\\
74.87 & 71.45 & 735$\pm$7 & 294$\pm$5 & 106$\pm$10 & 40$\pm$4
&1040$\pm$9\\
\hline
\hline
\end{tabular}
\end{center}
\end{table}

\begin{table} [t]
\caption{As in Table~\ref{yields-1} for the $^{11}$B~+~$^{209}$Bi
reaction. For the Rn and Fr isotopes only 15\% of the total yields
presented in the table contributes to the complete fusion
cross sections.}
\label{yields-2}
\begin{center}
\begin{tabular}{lllllll}
\hline \hline $E_{beam}$ \ & $E_{c.m.}$ \ & $\sigma_{fission}$  &
$\sum$$\sigma_{Ra}$ \ & $\sum$$\sigma_{Rn}$ \ &
$\sum$$\sigma_{Fr}$ \ \ & $\sigma_{CF}$\\
(MeV) \  & (MeV) \  & ~(mb)
 & ~(mb) \  & ~(mb) \ & ~(mb)
\ \ & (mb)\\
\hline
54.84 & 52.10 & 57$\pm$1 & 226$\pm$7 & ~~~-- & 1.9$\pm$1.0
& 283$\pm$7\\
59.87 & 56.88 & 176$\pm$2 & 403$\pm$7 & 11$\pm$6 & 1.3$\pm$1.0
& 581$\pm$7\\
64.85 & 61.61 & 329$\pm$3 & 462$\pm$6 & 38$\pm$4 & 6.4$\pm$0.8
& 798$\pm$7\\
69.87 & 66.38 & 521$\pm$5 & 471$\pm$7 & 58$\pm$6 & 15.4$\pm$2.0
& 1003$\pm$9\\
75.00 & 71.25 & 724$\pm$7 & 430$\pm$7 & 83$\pm$9 & 15.0$\pm$2.0
& 1169$\pm$10\\
\hline
\hline
\end{tabular}
\end{center}
\end{table}

\section{Suppression of Complete Fusion}

\begin{figure}[!]
\begin{center}
\includegraphics[width=8.00cm]{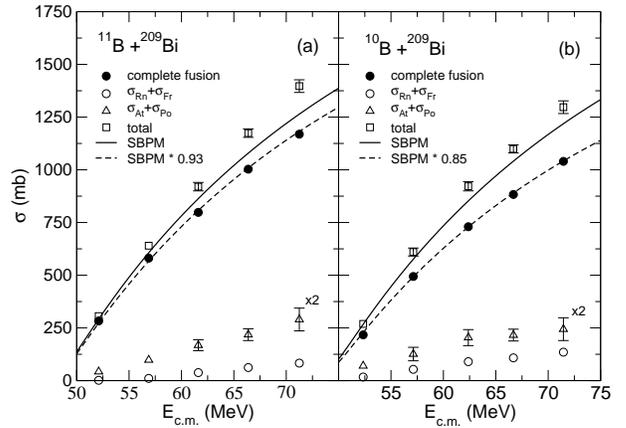}
\caption{Complete fusion excitation function (filled circles) for the
$^{10,11}$B~+~$^{209}$Bi reactions. The solid lines are the predictions of a
single barrier penetration model, and the dashed lines are the results of
these calculations multiplied by the indicated factor. Also shown are
the sum of the measured cross sections for Rn and Fr not associated with CF
(open circles), the sum of the measured cross sections for At and Po (open
triangles) and the total cross sections for almost all reaction products
heavier than the target (open squares). For clarity in the figure, the data
points plotted as open triangles have been multiplied by a factor of 2.}
\label{CF-2panels}
\end{center}
\end{figure}

To determine the suppression of fusion cross sections at
above-barrier energies due to breakup of the $^{10,11}$B
projectiles, SBPM calculations were performed using the S\~ao Paulo
potential (SPP) to give the real part of the nuclear potential~\cite{Ch02}.
In earlier works~\cite{Gas04,Cre05,Cre07}, the SPP has proved
reliable for reproducing fusion excitation functions for reactions
involving both strongly bound and weakly bound projectiles (CF+ICF),
and is particularly useful for measurements where experimental
fusion barrier distributions are not
available. The results of the SBPM calculations (solid lines) are
compared with the experimental CF
cross sections in Fig.~\ref{CF-2panels}. As expected, the measured
cross sections lie below the theoretical calculations.  The ratios
of the observed CF cross sections to those expected without breakup
(defined as F$_{\rm CF}$) are
F$_{\rm CF}$ = 0.85$\pm$0.01 for $^{10}$B and F$_{\rm CF}$ = 0.93$\pm$0.02 for
$^{11}$B. The larger reduction in CF
cross sections for the $^{10}$B induced reaction shows a correlation
with the lower breakup energy threshold for $^{10}$B (4.461 MeV)
compared with $^{11}$B (8.665 MeV), as might be expected.

A substantial part of the difference between the SBPM calculations
and the experimental CF cross sections ($\sim$ 70\%) can be found
in the measured yields of Fr and Rn isotopes which cannot be
associated with CF. The presence of these unexpectedly large yields
provides support for the conclusion from the SBPM calculations that
CF is suppressed in these reactions.

\subsection{Systematics of Fusion Suppression}

\begin{figure}[!]
\begin{center}
\includegraphics[width=7.00cm]{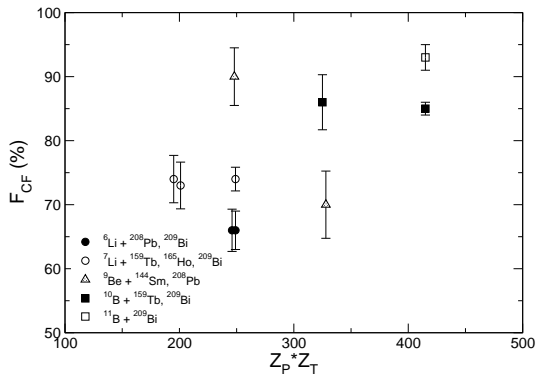}
\caption{The complete fusion suppression factor F$_{\rm CF}$ for reactions with
heavy target nuclei as a function of the charge product of projectile
and target.}
\label{SF}
\end{center}
\end{figure}

To investigate the systematics of the suppression of fusion due to
breakup, Fig.~\ref{SF} shows the CF suppression F$_{\rm CF}$ for reactions
involving weakly bound projectiles and heavy target nuclei. These
are for the reactions of $^{6,7}$Li incident on $^{159}$Tb~\cite{Muk06},
$^{165}$Ho~\cite{Tri02}, $^{208}$Pb~\cite{Wu03} and
$^{209}$Bi~\cite{Das04} targets, $^9$Be incident on $^{144}$Sm~\cite{Gom06}
and $^{208}$Pb~\cite{Das04} targets, and $^{10}$B
incident on $^{159}$Tb~\cite{Muk06}. The present result for
$^{10,11}$B~+~$^{209}$Bi are also shown. For the $^7$Li~+~$^{165}$Ho
reaction, the data~\cite{Tri02} have been reanalyzed to incorporate
changes in the CC calculations, including rotational couplings for
the $^7$Li as in Ref.~\cite{Das04}. This resulted in slightly less
suppression of CF (0.73) than the 0.70 reported in Ref.~\cite{Tri02}.

Understanding systematic behavior is important to clarify the
mechanism of break up and its effect on the fusion process. The
dependence of F$_{\rm CF}$ on the charge product of the reaction 
(Z$_{\rm P}$Z$_{\rm T}$) is illustrated in Fig.~\ref{SF}. The CF suppression for 
the reactions involving $^7$Li (open circles) and $^{10}$B (filled squares) 
projectiles are found to be rather independent of Z$_{\rm P}$Z$_{\rm T}$ within 
experimental error. The values for $^7$Li lie below those
for $^{10}$B, which may be expected due to the lower breakup energy threshold of
$^7$Li compared with $^{10}$B. For the measurements involving
$^9$Be, the CF suppression (open triangles in Fig.~\ref{SF}) shows a
strong variation with Z$_{\rm P}$Z$_{\rm T}$. The origin
of this difference is not clear. It may be related to the physical
mechanism of fusion suppression, or could possibly arise from
different experimental techniques used to obtain the CF excitation
functions.

\begin{figure}[!]
\begin{center}
\includegraphics[width=7.00cm]{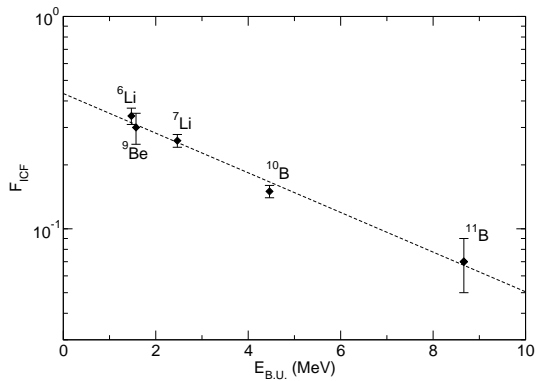}
\caption{The approximate fraction of ICF (F$_{\rm ICF}$) as a function of the
breakup threshold for reactions on $^{208}$Pb and $^{209}$Bi (see text). The
exponential line is to guide the eye.}
\label{Sup-Thr}
\end{center}
\end{figure}

This behavior makes it difficult to assess the dependence of the CF
suppression on breakup threshold for the full data set. However,
taking only the measurements on $^{208}$Pb and $^{209}$Bi targets,
the data are rather consistent. The reactions
$^6$Li~+~$^{209}$Bi~\cite{Das04} and $^6$Li~+~$^{208}$Pb~\cite{Wu03} give the
strongest suppression
(F$_{\rm CF}$ = 0.66). This is expected since the $^6$Li nucleus has the
lowest threshold against breakup (1.473 MeV). Assuming that the yield
of ICF is complementary to that of CF, the ICF fraction F$_{\rm ICF}$ is
given by F$_{\rm ICF}$ = 1 $-$ F$_{\rm CF}$; this is probably only approximately
true~\cite{Das04}. This ICF fraction is shown as a function of the projectile
breakup threshold in Fig.~\ref{Sup-Thr}. A remarkably consistent correlation is
seen, suggesting the dominant role of the breakup threshold. This
may suggest a too simple picture of breakup, as the probability of
excitation above the breakup threshold, as well as the threshold
energy itself, must play the crucial role. Measurements of breakup
at energies below the fusion barrier, where no significant capture
of the fragments is expected, and the energetics of the breakup
process may be reconstructed, are expected to give additional
insights into the physical mechanism(s) of
breakup~\cite{Tor07,Hin02}.

%%%%%%%%%%%%%%%%%%%%%%%%%%%%%%%%%%%%%%%%%%%%%%%%%%%%%%%%%%%%%%%%%%%%%%%%%%%%%%
\section{Summary and Conclusions}
\label{Summary}
%%%%%%%%%%%%%%%%%%%%%%%%%%%%%%%%%%%%%%%%%%%%%%%%%%%%%%%%%%%%%%%%%%%%%%%%%%%%%%

Above-barrier cross sections of  $\alpha$-active heavy reaction products, as 
well as fission, were measured for the reactions of $^{10,11}$B with 
$^{209}$Bi. The contributions of complete fusion and incomplete fusion in the 
evaporation residues were determined making use of data for the 
$^{30}$Si~+~$^{186}$W reaction, together with statistical model calculations. 
It was found that there are considerable contributions from ICF amongst the 
heavy reaction products. Fission was found to originate almost exclusively 
from CF, as determined by comparing experimental data for fusion reactions of 
$^4$He and $^{6,7}$Li with $^{209}$Bi. Compared with calculations of fusion 
without consideration of breakup (using the S\~{a}o Paulo nuclear potential) 
the CF cross sections are suppressed by 15~$\pm$~1\% ($^{10}$B) and 
7~$\pm$~2\% ($^{11}$B). Including these new results, for reactions involving 
light weakly bound stable nuclei bombarding targets of $^{208}$Pb and 
$^{209}$Bi, a remarkably consistent correlation of the suppression of CF 
(or equivalently the fraction of ICF) is found with the breakup threshold 
energy. 

Future planned measurements of below-barrier no-capture breakup, in
conjunction with the results presented here (and in previous papers)
of CF and ICF cross sections for reactions of projectiles with different 
breakup threshold energies, should be valuable in the development of 
realistic models describing breakup and fusion (complete and incomplete). In 
particular, in the near-future, a systematic study can test the recently 
developed three-dimensional classical dynamical model~\cite{Tor07}, which 
relates below-barrier no-capture breakup yields with above-barrier CF and ICF 
cross sections.

\begin{acknowledgments}
This work was supported by a Discovery Grant from the Australian
Research Council.
\end{acknowledgments}

%%%%%%%%%%%%%%%%%%%%%%%%%%%%%%%%%%%%%%%%%%%%%%%%%%%%%%%%%%%%%%%%%%%%%%%%%%%%%%
\appendix{
\section{Other Reaction Products}
\label{ICF}
%%%%%%%%%%%%%%%%%%%%%%%%%%%%%%%%%%%%%%%%%%%%%%%%%%%%%%%%%%%%%%%%%%%%%%%%%%%%%%

For the $^{10,11}$B induced reactions, ICF could result in the production of
many different isotopes of At and Po, as well as the Fr and Rn already 
discussed. For the At and Po isotopes, similar to the Rn isotopes, their direct
production cross sections are obtained by subtracting their feeding through 
Fr and Rn parent decays (see Fig.~\ref{production}). The direct production of 
At and Po isotopes can be associated with ICF and/or transfer reactions. The 
cross sections for production of both At and Po are given in
Table~\ref{At-ICF}, and shown in Fig.~\ref{At-xn} as a function of the
center-of-mass energy.

\subsection{At isotopes}

All Fr isotopes formed have $\alpha$-decay half lives short enough to
contribute to the prompt yield of At isotopes. However, the feeding of the 
observed At isotopes is rather small compared with their direct production, so
the feeding correction is not large. The observation of decay from the 
$^{212}$At isomeric state (spin=9$^-$; $T_{1/2}$=0.119 s) in the 
$\alpha$-spectra, which is not fed by the decay of $^{216}$Fr, confirms the 
direct production of this isotope.

\begin{figure}[t]
\begin{center}
\includegraphics[width=8.00cm]{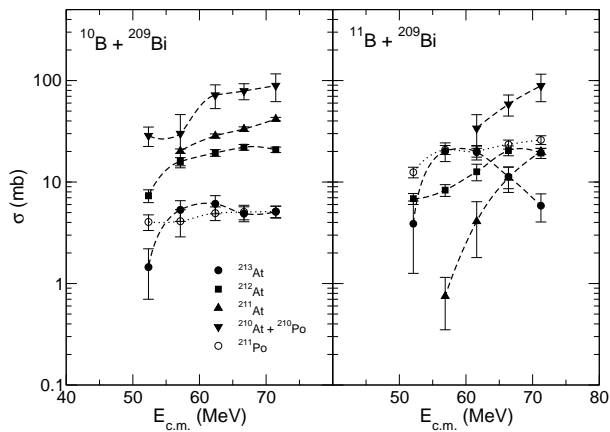}
\caption{The measured cross sections for At and Po
isotopes for the $^{10,11}$B-induced reactions. The
symbols denote the same products in the left and the right panels. The lines
 guide the eye.}
\label{At-xn}
\end{center}
\end{figure}

The cross sections for $^{210}$At could not be determined individually, since 
it decays almost completely (99.82\%) by electron capture to $^{210}$Po, with 
$T_{1/2}$=8.3 h. Assignment of yields to $\alpha$-transfer or ICF have not 
been attempted, as determination of the suppression of CF, the main aim of 
this work, is independent of this assignment.

\begin{table} [t]
\caption{The sum of the individual measured cross sections
for At and Po isotopes for the $^{10,11}$B reactions.
The quoted errors reflect only statistical uncertainties. The center-of-mass
energies $E_{c.m.}$ are corrected for energy losses in the target. The cross
sections for $^{209}$Po, which may be significant, could not be measured due
to its long half-life (102 years).}
\label{At-ICF}
\begin{center}
\begin{tabular}{llll}
\hline
\hline
$E_{beam}$ \ \ \   & $E_{c.m.}$ \ \ \
& $\sum$$\sigma_{At}$+$\sum$$\sigma_{Po}$ \ \ \
& $\sum$$\sigma_{At}$+$\sum$$\sigma_{Po}$\\
(MeV) \ \ \ & (MeV) \ \ \ & $^{10}$B~(mb) \ \ \ & $^{11}$B~(mb)\\
\hline
54.84 & 52.10 & 35$\pm$6   & 21.3$\pm$3.4\\
59.87 & 56.88 & 63$\pm$16  & 49$\pm$5\\
64.85 & 61.61 & 102$\pm$19    & 84$\pm$13\\
69.87 & 66.38 & 108$\pm$14     & 109$\pm$14\\
75.00 & 71.25 & 122$\pm$27     & 145$\pm$27\\
\hline
\hline
\end{tabular}
\end{center}
\end{table}

\subsection{Po isotopes}

The total cross sections for production of $^{210}$Po ($T_{1/2}$=138 days) were 
obtained from off-line spectra accumulated over a few days after the 
irradiations, when all $^{210}$At will have decayed to $^{210}$Po. A significant
fraction of the $^{210}$Po yield may be due to transfer of one proton to 
$^{209}$Bi,  with ground-state $Q$-values of -1.60 MeV ($^{10}$B~+~$^{209}$Bi) 
and -6.24 MeV ($^{11}$B~+~$^{209}$Bi). Deuteron transfer reactions are also 
likely to populate $^{211}$Po. The summed $^{211}$Po ground-state and $^{211}$Po
isomeric state [spin=(25$^+$/2);$T_{1/2}$=25.2 s] yields are shown as open 
circles in Fig.~\ref{At-xn}. The population of the high spin isomeric state 
implies that ICF is a significant route of population. The cross sections for 
$^{209}$Po could not be measured due to its long half-life of 102 years. Based 
on the sizable production of $^{210}$Po compared with $^{211}$Po, a substantial 
yield of $^{209}$Po might be expected. Thus in this work we could only 
determine the lower limit of the total cross section of At and Po.

\end{document}